\documentclass[12pt]{article}
\usepackage{natbib}
\usepackage{tikz}
\usepackage{bm}
\usepackage{amsmath}

\newcommand{\bra}[1]{\langle#1|}
\newcommand{\ket}[1]{|#1\rangle}

\begin{document}
\setcitestyle{semicolon,aysep={}}
\date{10 November 2022}
\title{Identical Particles in Quantum Mechanics: Against the Received View \thanks{Draft of a chapter in \textit{Individuals and Non-Individuals in Quantum Mechanics}, to appear with Springer; second version.}}
\author{Dennis Dieks \\History and Philosophy of Science\\
Utrecht University}
\maketitle
\begin{abstract}
According to the ``Received View''  identical quantum particles are a previously unknown kind of objects that do not possess individuality. In this Chapter we discuss this view, criticize it, and propose an alternative. According to this alternative view so-called ``identical quantum particles''  should in many cases not be seen as objects---particles---at all. However, there are situations in which a particle picture does become applicable. But the particles that emerge in these cases are distinguishable individuals, unlike the particles of the Received View. 
 \end{abstract}

\section{Introduction}

From its first beginnings quantum mechanics has been beset by problems concerning the nature of the objects it deals with. Controversy started already soon after \cite{planck} had introduced discrete energy packets $h\nu$ in his theoretical derivation of the law of black body radiation (which determines the distribution of energy over the frequencies of electromagnetic radiation, in thermal equilibrium). 
Planck had set out to calculate the distribution of energy over  resonators in the wall of a box containing radiation, because such resonators emit and absorb radiation, thus establishing thermal equilibrium with the electromagnetic field. The way the energy is distributed over the resonators therefore provides information about the energy distribution in the electromagnetic field. For his calculation Planck needed the entropy of the resonators, which he determined by using the celebrated Boltzmann expression $S=k\log W$, with $W$ the number of ways a given amount of energy can be distributed over $N$ resonators. It is difficult to get a grip on this number, since energy is a continuous quantity.  Planck therefore reduced the problem to the more tractable one of distributing $P$ energy packets over $N$ resonators, by making use of the ``mathematical trick'' of discretizing the energy (making continuous quantities discrete in order to simplify computations is a standard ploy in mathematical physics).  To solve the ensuing combinatorial problem, Planck treated the resonators as fixed and distinct physical individuals, but the energy packets as quantities without physical identity. That is, he supposed that the interchange of two or more energy packets would not lead to a new configuration. 

The intuition behind this difference in treatment of resonators and energy packets seems clear.  The resonators are objects that differ in one or more of their physical attributes: at least their fixed positions in the walls of the container distinguish them from each other. By contrast, it is unclear what it even means to replace one part of a total amount of energy by another part. 
Classically, the energy possessed by a physical system is a continuous total quantity, and there is no natural way of subdividing it into individual parts with their own characteristics.

One might object that in Planck's calculation energy is quantized: Planck assumed that the resonators can only possess energy values $n.\epsilon$, with $n$ an integer and $\epsilon$ the energy contained in one packet. Doesn't this make an essential difference for the question of whether it makes sense to switch energy elements? However, quantization of the total energy does not automatically imply the independent existence of individual energy units. Think of a jug, always filled with a quantity of liquid that is an integer multiple of deciliters, because the filling mechanism can deliver only discrete volumes. In spite of the fact that the amount of liquid is ``quantized'', there is no first, second, third, etc., deciliter in the jug once it has been filled.  

Against this analogy one might argue that in a microscopic description there do exist individual constituents of the fluid (classical molecules or atoms) and that in principle it would be possible to follow these during the filling process. In this way one could define the first deciliter that came in, and so on. Note, however, that this counterargument does not show that the quantized nature of the volumes entails that these volumes are  composed of individual units; the argument starts from the premise that there exist individual fluid constituents, quite independently of the volume quantization,  and its conclusion is independent of whether the total fluid amount is quantized. 

An alternative analogy, first proposed by \cite{schrod}, avoids the objection about atoms and molecules and captures the situation Planck was dealing with better. 
Schr\"odinger discusses the case of money in a bank account---scriptural (deposit) money, not physical coins and banknotes in a safe. The total amount of money in the account is always an integer times the monetary unit---euros, say. Suppose the balance is 100 euros; does it make sense to ask for the first euro? Clearly not: there are no individual euros in the account. Suppose that I take a 1 euro coin to a bank branch and deposit it in my account. The balance will change to 101 euros, but it is meaningless to pose the question which of these 101 euros corresponds to my coin: there is no hundred-and-first euro.  

Does this mean that we should think that the euros in the account are entities of a previously unknown kind, namely entities that lack individuality? Entities that are fundamentally indistinguishable from each other, and to which the notion of identity does not apply? In order to consistently discuss such putative identity-less objects we should change the usual logic and mathematics that we use. Indeed, in standard set theory each member of a set automatically has its own identity (namely the property represented by the singleton set of which it is the sole member).  However, introducing non-standard set theory in order to discuss bank accounts seems  an extravagant decision. After all, there is nothing mysterious or unsatisfactory in the usual description of situations of the bank account type: both bankers and clients only need standard logic and set theory. Common-sense simply dictates that there \emph{are no} independently existing parts of the total buying power of a bank account. 

It is true that under certain conditions amounts of money may be said to acquire an identity. Withdrawn amounts can be told apart after they have landed in different, previously empty, bank accounts, and similarly flows of money may sometimes be traced over time. In a scenario in which a collection of bank accounts can each contain only either $1$ or $0$ euro, one might maintain that the euros can be identified individually by the account they are in (see  \citep{die4}, also for comparison with the quantum case).  This is because the euros ``occupy different states'', namely different accounts, identifiable by their account numbers, owners, etc.

In the following we will discuss and defend a way of dealing with ``identical quantum particles'' that is similar to the story just sketched for bank money. That is, we will consider what is usually called a collection of ``identical particles that are in the same state'' as \emph{one} object,  not consisting of individual parts; this quantum object is identified by the quantum state and its occupation number (the analogy is with a bank account and its balance). Under certain circumstances it may happen, though, that distinguishing characteristics are created (analogously to what may happen in bank transfers) and that the notion of an individual quantum particle becomes applicable.  

By going this way we will deviate from what in the philosophy of physics has become known as the ``Received View'' regarding the nature of identical quantum particles \citep{french}. The Received View was to a large extent motivated by developments that took place after Planck's introduction of discrete energy packets. First there was Einstein's suggestion that black body radiation itself might be considered to consist of ``light quanta'', behaving as particles, whereas Planck had only spoken about quantized energy exchanges with material resonators \citep{einstein}. Einstein's suggestion remained controversial, but started to gain acceptance when two decades later  \cite{einstein2} showed that Planck's statistical treatment of energy packets could also be applied successfully to the atoms of a quantum gas.  The permutation invariance typical of Planck's statistics remained in place in this new application---but now Planck's energy packets were replaced by gas atoms, i.e.\ \emph{quantum particles}. Apparently the exchange of such particles does not correspond to any physical  difference. This leads to the core idea of the Received View: quantum particles of the same kind are objects of a completely novel type, namely objects without identity.                

 \section{The Received View}

\subsection{Quantum statistics}\label{ehrenfest}
Planck's derivation of the black body radiation law hinged on the calculation of the entropy of a collection of resonators that exchange energy with the radiation in a container. If this energy exchange can only take place in discrete amounts, namely  multiples of $\epsilon = h\nu$, with $\nu$ the radiation's frequency, the total energy of the $N$ resonators is $P\epsilon$, and the entropy is proportional to the logarithm of the number of ways  this total energy can be distributed over the resonators.   \cite{planck} had been brief about how to determine this number: he had simply written down the answer 
\begin{equation}\label{boseinst}
C^{N}_{P} = \frac{(N-1+P)!}{(N-1)!P!},
\end{equation}
referring to a mathematics text containing a somewhat similar problem.  This did not satisfy \cite{ehrenfestkam}, who set out to provide a deduction of formula (\ref{boseinst}) in which the physical premises would be perspicuous. In order to do so, they represented the possible states of the resonator system by ``symbols''; for the case in which resonator $1$ possesses the energy $4\epsilon$, oscillator $2$ the energy $2\epsilon$, oscillator 3 $0\epsilon$ (no energy) and oscillator $4$ the energy $\epsilon$, this representative  symbol takes the form \citep[870-871]{ehrenfestkam}: 
\begin{center}
\begin{tikzpicture}
\draw[very thick] (0,-0.5) -- (0,0.5); 
\draw[very thick] (0.1,-0.5) -- (0.1,0.5); 
\draw[very thick] (8,-0.5) -- (8,0.5);
\draw[very thick] (8.1,-0.5) -- (8.1,0.5);
\draw[very thick] (3.5,0) circle(0.2cm);
\draw[very thick] (5.6,0) circle(0.2cm);
\draw[very thick] (6.5,0) circle(0.2cm);
\node at (0.7,0){$\bm{\epsilon}$};
\node at (1.4,0){$\bm{\epsilon}$};
\node at (2.1,0){$\bm{\epsilon}$};
\node at (2.8,0){$\bm{\epsilon}$};
\node at (4.2,0){$\bm{\epsilon}$};
\node at (4.9,0){$\bm{\epsilon}$};
\node at (7.3,0){$\bm{\epsilon}$};
\end{tikzpicture}
\end{center}

The small circles indicate boundaries between the individual resonators; the resonators themselves are ordered from left to right and could be given individual labels or names. The background of formula (\ref{boseinst}) now becomes clearer: the number of possible energy distributions, given fixed values of  $P$ and $N$, equals the number of different symbols of this kind that can be written down with those values of $P$ and $N$. In each such symbol there are $P$ instances of $\bm{\epsilon}$ and $(N-1)$ instances of the sign   \begin{tikzpicture}
\draw[very thick] (0,0) circle(0.2cm);
\end{tikzpicture}.  As \cite{ehrenfestkam} write:
\begin{quote}
first considering the $(N-1 + P)$ elements $\bm{\epsilon}$...$\bm{\epsilon}$, \begin{tikzpicture}
\draw[very thick] (0,0) circle(0.2cm);
\end{tikzpicture}...\begin{tikzpicture}
\draw[very thick] (0,0) circle(0.2cm);
\end{tikzpicture}
as so many distinguishable entities, they may be arranged in 
\begin{equation}\label{2}
(N - 1 + P)!\end{equation}
different manners between the ends. Next note, that each time 
\begin{equation}\label{3}
(N-1)! P! \end{equation} of the combinations thus obtained give the same symbol for the distribution (and give the same energy-grade to each resonator), viz.\ all those combinations which are formed from each other by the
permutation of the $P$ elements $\bm{\epsilon}$ or the $(N -1)$ elements \begin{tikzpicture}
\draw[very thick] (0,0) circle(0.2cm);
\end{tikzpicture}. The number of the \emph{different} symbols for the distribution and that of the distributions themselves required is thus obtained by dividing (2) by
(3). q.e.d.
\end{quote}

The essential premise of the derivation is therefore the permutability of the signs \begin{tikzpicture}
\draw[very thick] (0,0) circle(0.2cm);
\end{tikzpicture} among each other,  and similarly for the signs $\bm{\epsilon}$. The former permutability is self-evident: the small circles are purely formal devices introduced for the purpose of demarcating the individual resonators from each other; switching two of these signs does not correspond to any physical change. The permutability of the energy elements  $\bm{\epsilon}$ is more problematic. One might be tempted to think that these signs refer to ``light particles'',  especially in light of Einstein's light quantum hypothesis \citep{einstein}. If that interpretation of the $\bm{\epsilon}$ signs is adopted, it is not evident that there can be no differences between situations differing by the permutation of a number of $\epsilon$-s; in the Boltzmannian approach particle permutations are standardly taken to lead to new configurations.  

\cite{ehrenfestkam} devote an Appendix to this issue, entitled \emph{The contrast between Panck's hypothesis of the energy-grades and Einstein's hypothesis of energy-quanta} (this appendix is actually longer than the main text of their paper).  They emphasize and warn:
\begin{quote}
The permutation of the elements $\bm{\epsilon}$ is a purely formal device, just as the permutation of the elements \begin{tikzpicture}
\draw[very thick] (0,0) circle(0.2cm);
\end{tikzpicture} is. More than once the analogous, equally formal device
used by Planck, viz.\ distribution of $P$ energy-elements over $N$ resonators, has by
a misunderstanding been given a physical interpretation [...]  

Planck does not deal with really mutually free
quanta $\bm{\epsilon}$; the resolution of the multiples of $\bm{\epsilon}$ in separate
elements $\bm{\epsilon}$, which is essential in his method, and the introduction
of these separate elements have to be taken \emph{cum grano salis}; it is simply a formal device entirely analogous to our permutation of the elements $\bm{\epsilon}$ or \begin{tikzpicture}
\draw[very thick] (0,0) circle(0.2cm);
\end{tikzpicture}. The real \emph{object which is counted} remains the number of all the different distributions of $N$ resonators over the energy-grades $0, \bm{\epsilon}, 2\bm{\epsilon},... $ with a given total energy $P \bm{\epsilon}$.
\end{quote}

\cite{ehrenfestkam} thus maintain that the energy elements should not be thought of as physical entities but rather as  merely formal devices for handling the quantization of energy. They back up this claim, in the remainder of their text, by pointing out that \cite{einstein} had assumed his light quanta to be statistically independent individuals, and that this had led to Wien's radiation law. But Wien's law is incorrect;  Planck's law is the empirically correct one. So in order to obtain empirically correct predictions we must renounce the idea that the $\epsilon$-s stand for individual light particles. 

Planck's law requires a probability distribution in which the number of equiprobable cases is given by Eq.\ (\ref{boseinst})---what we now call the Bose-Einstein distribution. The difference with the usual particle statistics   is illustrated by Ehrenfest and Kamerling-Onnes with the help of an example in which three energy elements have to be distributed over two resonators: according to Einstein's 1905 line of reasoning this can be done in $2^3=8$ ways (each of the three light quanta has an independent choice between two resonators), whereas formula (\ref{boseinst}) tells us that there are only four ways of doing this. The difference is due to the fact that according to Eq.\ (\ref{boseinst}) only situations in which the total resonator energies are different count as distinct, whereas according to Einstein's original argument it would also make a difference \emph{which} energy element is in \emph{which} resonator. \cite{ehrenfestkam} therefore conclude: ``Planck's \emph{formal device} (distribution of P energy-elements $P \bm{\epsilon}$ over $N$ resonators) \emph{cannot be interpreted in the sense of Einstein's light-quanta}.'' In other words, the energy elements should not be thought of as particles.

\subsection{Particles without identity}\label{noidentity}
The situation changed drastically when \cite{bose} applied the statistics of Eq.\ (\ref{boseinst}) to a ``gas'' consisting of light quanta, without any consideration of energy exchanges with resonators, and was thus able to derive Planck's law; and when \cite{einstein2} generalized this idea by using the same statistics to calculate the entropy of a mono-atomic ideal quantum gas. These developments made the distribution (\ref{boseinst}) into a distribution of \emph{particles}, despite Ehrenfest's and Kamerling-Onnes'  qualms about the lack of statistical independence between such particles. The essential novel feature is that permutations between particles, even when they are in different states (as in the permutation from ``particle 1 in state A and particle 2 in state B'' to ``particle 2 in state A and particle 1 in state B''),  do not give rise to new configurations. 

This change of perspective, from Bose-Einstein statistics as a formal device to its role as the statistics of quantum particles, was consolidated by the development of modern quantum mechanics, with its symmetrization rules for particles of the same kind (``identical particles'').
Suppose, in analogy with the example discussed by Ehrenfest and Kamerlingh-Onnes,  that we have three identical quantum particles that each can be in one of two pure quantum states, $| A \rangle $ and $|B \rangle $. In this case at least two particles must occupy the same state, so the particles have to be bosons and  the state must be symmetric according to the symmetrization rules.  Quantum mechanics assigns a state to this three-particle system in the tensor product Hilbert space ${\cal{H}}_1 \otimes {\cal{H}}_2 \otimes {\cal{H}}_3$, where the factor spaces ${\cal{H}}_1, {\cal{H}}_2, {\cal{H}}_3$ are one-particle Hilbert spaces. The standard interpretation is that ${\cal{H}}_i$ is the state space of particle $i$, so that the labels of the three factor spaces also label the particles. 

There are now four possible states:
\begin{align}
|\Psi \rangle_1  & = | A \rangle_1 |A \rangle_2 | A \rangle_3  \\
|\Psi \rangle_2  & = | B \rangle_1 |B \rangle_2 | B \rangle_3   \\
|\Psi \rangle_3  & = \frac{1}{\sqrt{3}} \{ | B \rangle_1 |A \rangle_2 | A \rangle_3  +  | A \rangle_1 |B\rangle_2 | A \rangle_3  + | A \rangle_1 |A \rangle_2 | B \rangle_3 \} \\
|\Psi \rangle_4   & = \frac{1}{\sqrt{3}} \{ | A \rangle_1 |B \rangle_2 | B \rangle_3  + | B \rangle_1 |A \rangle_2 | B \rangle_3 + | B \rangle_1 |B \rangle_2 | A \rangle_3  \}
\end{align}

That there are only these four possibilities is analogous to Ehrenfest's and Kamerling-Onnes' example in which an undifferentiated amount $3 \bm{\epsilon}$ of quantized energy was distributed over two resonators. In that case all energy could be possessed by resonator $A$, or all could be possessed by  $B$; or $A$ could have the energy $2\bm{\epsilon} $ and $B$ the energy $\bm{\epsilon}$, and \emph{vice versa}.  In the statistical calculations needed to determine thermodynamic quantities these four possible energy configurations were assigned equal weights.   

In the statistical mechanics of many-particle quantum systems  the states (4)--(7) are similarly assigned equal probabilities. But now we are dealing with descriptions of particles instead of energy quanta. The fact that states (6) and (7) receive the same statistical weights as (4) and (5),  suggests that particle permutations do not correspond to physical differences.  This suggestion motivates what French and Krause have baptized the \emph{Received View}. As \cite{french} write (p.\ 143):
\begin{quote}
from the point of view of the statistics, the particle labels are otiose. The implication, then, is that the particles can no longer be considered to be individuals, that they are, in some sense, `non-individuals'. This conclusion expresses what we have called the `Received View': classical particles are individuals but quantum particles are not. ... As we shall see in the rest of the book, one can in fact go beyond mere metaphor and underpin the Received View with an appropriate logico-mathematical framework.
\end{quote}

The formal framework here referred to is that of ``quasi-set theory'' \citep[Ch.\ 7]{french}. In this variation on standard  set theory two sorts of ``atoms'' (``\emph{Urelemente}'') are admitted, via the introduction of predicates $m(x)$ and $M(x)$ signifying that $x$ is an $m$-atom or an $M$-atom, respectively.  The intended interpretation is that $m$-atoms will refer to quantum particles, whereas $M$-atoms refer to classical objects. Because of the motivating idea that quantum particles do not possess individuality, the domain of application of the concept of identity ($=$) in quasi-set theory is restricted to $M$-atoms. There \emph{is} a notion of identity in quasi-set theory, but its applicability  is limited to (quasi-)sets and $M$-objects, so that $x=y$ is not a well-formed formula if at least one of the elements $x$ and $y$ is an $m$-atom. There is a relation of ``indistinguishability'' (``$\equiv $''), which is weaker than identity and is applicable to $m$-atoms. A quasi-set all of whose members are indistinguishable $m$-atoms will have a ``quasi-cardinal'' indicating the total number of elements of the quasi-set, but not an ordinal, since the atoms cannot be labeled (labeling would provide individual names and thus bestow individuality on the $m$-atoms). Without going into further technical details (see \citep{french,arenhkrause,arenhart}), we might say that in this quasi-set theory $m$-atoms are handled asif they were  objects, in the sense that there are variables for them, that it is possible to quantify over them, and that they may differ  in the sense of not being elements of the same set---as in ordinary set theory. Nevertheless, certain arguments valid in ordinary set theory are blocked for $m$-atoms because the notion of identity does not apply to them.  

For example, two $m$-atoms may be indistinguishable, $x \equiv y$, sharing all properties, without being the same. In ordinary set theory this is impossible: one of the shared properties would correspond to ``being identical to $x$'', since $x$ is identical to itself, $x = x$. But in quasi-set theory the notion of self-identity is not applicable to $m$-atoms, so that Leibniz's principle does not hold. By the same token, the singleton set ${a}$ cannot be formed if $a$ is an $m$-atom: this set would consist of all atoms that are identical to $a$, but the notion of identity is not defined for $m$-atoms.  

This axiomatic quasi-set theory is meant to capture the nature of quantum particles as entities without identity and to provide a formal background to the Received View. For example, permutations of $m$-atoms lead to indistinguishable situations according to quasi-set theory, which represents ``one of the most basic facts regarding indistinguishable quanta'' \citep{domenech2}. The theorem in quasi-set theory which formalizes this indistinguishability of permuted configurations reads \citep[p.\ 3086]{domenech2}:
\begin{quote}
Let $x$ be a finite quasi-set such that $x$ does not contain all elements
indistinguishable from $z$, where $z$ is an $m$-atom such that $z
\in x$. If $w \equiv z$ and $w \notin x$, then there exists $w'$
such that $(x - z') \cup w' \equiv x$.
\end{quote}
Here $z'$ and $w'$ stand for quasi-sets with quasi-cardinal $1$ whose
only elements are indistinguishable from $z$ and $w$, respectively. The idea is that if an element of a quasi-set is replaced by an element that is indistinguishable from it, but originally was not in the same quasi-set, the final situation cannot be distinguished from the original one. 

\subsection{Quasi-sets of quantum particles}\label{quasiparticles}

If the labels that are standardly employed in many-particle quantum mechanics turn out to be otiose in the description of particles of the same sort, it should be possible to construct a formalism that does without particle labels. The task to reconstruct the quantum mechanics of many-particle systems accordingly was undertaken by \cite{domenech1}. Their  \emph{ab initio} label-less quantum theory of particles without identity takes the following form.

Consider a set of eigenvalues of a quantum mechanical quantity (an ``observable''); for the sake of concreteness take the eigenvalues of the Hamiltonian  $H$ of the system, so that
$H|\varphi_{i}\rangle=\epsilon_{i}|\varphi_{i}\rangle$, with $|\varphi_{i}\rangle$ the energy eigenstates. Now introduce the notion of a ``quasi-function''; this is a mapping associating a finite quasi-set with each value $\epsilon_i$, so that disjoint quasi-sets are associated with different $\epsilon$-values. It is assumed that the sum of the
quasi-cardinals of the quasi-sets occurring in this mapping is finite. If the quasi-set $x$ is associated with $\epsilon_{i}$,  the interpretation is that the energy level
$\epsilon_{i}$ has the quasi-cardinal $qc(x)$ as its occupation number. An alternative way of representing the situation is with symbols like 
$f_{\bm{\epsilon_1}\bm{\epsilon_1}\bm{\epsilon_1}\bm{\epsilon_1}\bm{\epsilon_2}\bm{\epsilon_2}\bm{\epsilon_4}}$, meaning that the level $\epsilon_{1}$ has occupation number $4$ while
the levels $\epsilon_{2}$ and $\epsilon_{4}$ have the occupation numbers 
$2$ and $1$, respectively. The levels that do not appear are understood to have occupation number zero---note the similarity to the ``symbols'' of Ehrenfest and Kamerling-Onnes discussed in section \ref{ehrenfest}. 

The use of pure quasi-sets (\emph{i.e.}, quasi-sets solely consisting of $m$-atoms) makes the use of particle labels not only superfluous but impossible, since quasi-sets cannot be ordered. As \cite{domenech1} remark, ``the only reference is to the occupation numbers, because permutations make no sense here, as it should be.'' 

Anticipating our own analysis of the situation, we would like to comment that the latter remark hits the nail on its head: if labels cannot be defined, permutations as ordinarily defined make no sense. It would seem natural to take this as a signal that there are no physical grounds for assuming the existence of a multiplicity of objects populating the various states. This move, however, would eliminate the motivation for the introduction of quasi-sets of quantum objects occupying a state. We will say more about this worry in the next section.

As \cite{domenech1} show, state descriptions of the form $f_{\bm{\epsilon_1}\bm{\epsilon_1}\bm{\epsilon_1}\bm{\epsilon_1}\bm{\epsilon_2}\bm{\epsilon_2}\bm{\epsilon_4}}$, in which quasi-cardinalities of quasi-sets denote the occupation numbers of eigenstates like  $|\varphi_{i}\rangle$, can be taken as building blocks (by converting them into vectors constituting a basis) for the construction of a many-particle Hilbert space. Not surprisingly, the resulting formalism is identical to the Fock space formalism of quantum field theory, which in turn implies that the constructed Hilbert space is isomorphic to either the symmetric (bosons) or antisymmetric (fermions) sector of the usual tensor product many-particle Hilbert space\footnote{A rival method for handling identical particles without the introduction of labels was proposed by Lo Franco and co-workers \citep{lofranco1,lofranco2,dieks10}. This approach also boils down to the Fock space formalism of quantum field theory.}.

The Received View thus comes in two flavors. First, there is the labeled tensor product Hilbert space formalism. In this formalism labels are assumed to refer to particles, but permutations of labeled particles are declared to not represent any physical differences. Second, there is the quasi-set formalism in which labeling is ruled out from the outset. Still, the existence is accepted of separate entities that can be permuted in the quasi-set theoretical sense explained in section \ref{noidentity}. These entities do not possess any individuating physical properties---they must all share the same attributes, just as in the labeled formalism. One might say, a bit ironically, that the quasi-set version of the Received View is the labeled version plus the refinement that all labels have been removed because labeling is impossible. 

\section{Criticism of the Received View}\label{criticism}

The history of the Received View shows how the notion of particles without identity originated in the application to particles of statistics devised for cases in which there is an undifferentiated whole.  This amalgam of different conceptual frameworks suffers from internal tensions: an undifferentiated whole does not consist of non-arbitrary parts, whereas particles as traditionally conceived are paradigmatic individual objects. 

The particles of classical physics exemplify the latter point: they are impenetrable localized entities that travel along well-defined spatial paths.  They can consequently be distinguished from each other at any instant of time and can be followed over time (they possess ``genidentity''). This makes it possible to label classical particles in a physically meaningful way: their labels can be associated with identifying physical properties (at least position, possibly also other properties like mass, charge etc.). 

The discussions surrounding the status of identical quantum particles may suggest that in the practice of quantum mechanics the use of the concept ``particle'' is completely different. It then may come as a surprise that paradigm cases of the use of ``particle'' in experimental quantum physics are very much like the examples from classical physics. Individual elementary particles can be identified by their paths in devices like a bubble chamber; single electrons can be trapped in a potential well and kept there for a long time; an electron gun can be set to fire one single electron, which a bit later hits a detector; and so on.  In such laboratory situations the attribution of identity to quantum particles is a matter of course.  

Categorizing the world in terms of objects that can be handled separately is a natural part of physical methodology. This conceptual framework motivates even the form of quasi-set theory: although the notion of identity has been removed, quasi-set theory is still about ``atoms''  of which there are definite numbers, which can belong to different quasi-sets, and thus differ from each other. However, there is a tension here: as soon as we commit to things that exist in definite numbers and can differ, the notion of identity is looming large.  As \cite{berto} argues (see also \citep{jantzen,dorato,bueno}):    
\begin{quote}
That a sentence of the form $a = b$  is true, ..., means that we need to count one thing: the
thing named $a$, which happens to be the thing named $b$... That we, instead, count two things, means that that sentence is false. But then its negation, $\neg (a = b)$, is true. So $a$ and $b$ are different. And if the concept of difference meaningfully applies to $a$ and $b$,
the one of identity does as well. $a = b$ is meaningful together with its negation:
adding or removing a negation in front of such a meaningful sentence cannot
turn it into a meaningless one. The concept of identity cannot but apply to
whatever the concept of difference applies to: if---to use Ryle’s jargon---we have
no category mistake in the latter case, we have no such mistake in the former.
When the number of things (in a system) is given by positive integer $n$, these
things cannot lack self-identity.
\end{quote}

There is therefore a tension between the notions of separate atoms on the one hand and the absence of identity of these atoms on the other. But apart from the conceptual confusion apt to result from this tension, there is also a methodological reason to be wary of the notion of identity-less objects. In empirical sciences like physics we need a justification, in the final analysis based on empirical data, for making distinctions. In our case, we need a convincing empirical motivation for speaking about separate particles at all in situations where we also have to acknowledge that these putative particles lack identity. If it is accepted that no demarcation lines between the proposed particles can be drawn, not even in theoretical considerations, there appear to be no grounds for conceptualizing the situation in terms of constituent particles rather than wholes. 

The analogy of deposit money in a bank account again illustrates the predicament: there is nothing in the total sum of money in an account suggesting the independent existence of entities that together build up the account's balance. To discuss bank accounts in terms of  quasi-sets of monetary units lacking identity, with  quasi-cardinalities, does not illuminate the nature of bank accounts---it instead obscures the fact that only total buying power is significant. 

It is exactly this problematic introduction of identity-less entities for the purpose of discussing what can unproblematically be considered as undifferentiated wholes that lies at the heart of the Received View.  As we have seen in section \ref{quasiparticles}, the Received View\footnote{In the form elaborated by \cite{domenech1} and \cite{domenech2}.} associates different quantum states (to be compared with different bank accounts) not only with occupation numbers (comparable to total account values), but also with quasi-sets whose quasi-cardinals equal the occupation numbers. That is, instead of saying that state $|\varphi_{i}\rangle$ has occupation number $n_i$, so that there is an energy $n_i. \epsilon_{i}$ in the ``mode'' represented by $|\varphi_{i}\rangle$, the Received View holds that there are $n_i$ separate though identity-less particles in the state  $|\varphi_{i}\rangle$.  

These quantum particles without identity cannot be ordered and labeled by natural numbers  and therefore cannot be counted in the ordinary sense of the word.\footnote{Adherents of the Received View retort that  counting identical quantum particles need not involve a mapping to the natural numbers. For example, \cite{krausearenh} state that there exist alternative counting procedures, like the \emph{weighing} of a total amount, that are able to determine numbers of identity-less entities. Quasi-cardinalities can similarly be determined by measuring the total amount of energy in mode  $|\varphi_{i}\rangle$. This is analogous to arguing that the euros in a bank account \emph{can} be counted, namely by looking at the value of the account. But clearly, the existence of a total account value cannot be regarded as support for the actual existence of quasi-sets of euros in an account; nor can the existence of an occupation number be seen as support for the existence of several quasi-particles  in state $|\varphi_{i}\rangle$. Quite the opposite: in such cases the non-existence of a counting procedure in the ordinary sense, \emph{i.e.}\ a mapping to the natural numbers, disconfirms the existence of building blocks that constitute the whole.}  Thinking in this way about particles is far removed from physical practice. As already pointed out, the laboratory practice of quantum physics is replete with talk about particles, but these particles are countable in the ordinary sense, possess identity, and behave more and more like classical particles when the classical limit of quantum mechanics is approached. The same particle concept is standard in theoretical considerations. Think, for example, of the  well-known arguments concerning how the uncertainty relations make quantum particles different from their classical counterparts. The uncertainties  in position and momentum of a quantum particle obey a relation of the form  $\Delta Q. \Delta P \geq \hbar$, so that a very small value of $\Delta Q $ implies a large value of $\Delta P $.  Since $P=mv$, with $m$ the mass and $v$ the speed of the particle, we can nevertheless have a very small uncertainty in $v$  if the particle is macroscopic (with a very large mass).  This is standardly taken to imply that a single individual quantum particle to a high degree of approximation will have a well-defined trajectory and will become indistinguishable from a classical particle in the macroscopic limit.  It follows that at least in these cases particles in the quantum regime are conceived of as identifiable and distinct from their fellow particles of the same sort---and therefore can be assigned an identity.  There thus proves to be a mismatch between the quantum particles allowed by the Received View, which never possess identity, and the objects called particles in the actual practice of quantum physics.  

\section{Identical Particles as Distinguishable Objects}\label{QP}

The discrepancy between the particle notion used in physical practice and the one of the Received View is highlighted in  the warning issued by \citeauthor{domenech1} (2008, p.\ 974) when they explain the Received View:
\begin{quote}
before to continue we would like to make some few remarks on a common misunderstanding... People generally think that spatio-temporal location is a sufficient condition for individuality. Thus, two electrons in different locations \emph{are} discernible, hence \emph{distinct individuals}...  we recall that even in quantum physics, fermions obey the Pauli Exclusion Principle, which says that two fermions (yes, they `count'
as more than one) cannot have all their quantum numbers (or `properties') in common. Two electrons (which are fermions), one in the South Pole and another one in the North Pole, \emph{are not individuals in the standard sense} (and we can do that without
discussing the concepts of space and time). Here, by an individual we understand an object that obeys the classical theory of identity of classical (first or higher order) logic (extensional set theory included). In fact, we can say that the electron in the South Pole
is described by the wave function $\psi_S(x)$, while the another one
is described by $\psi_N(x)$ (words like `another' in the preceding
phrase are just ways of speech, done in the informal metalanguage).
But the wave function of the joint system is given by
$\psi_{SN}(x_{1},x_{2})=\psi_S(x_{1})\psi_N(x_{2})-
\psi_N(x_{1})\psi_S(x_{2})$ (the function must be anti-symmetric in
the case of fermions, that is, $\psi_{NS}(x_{1},x_{2})= -
\psi_{NS}(x_{2},x_{1})$), a superposition of the product wave
functions $\psi_S(x_{1})\psi_N(x_{2})$ and
$\psi_S(x_{2})\psi_N(x_{1})$. Such a superposition cannot be
factorized. Furthermore, in the quantum formalism, the important
thing is the square of the wave function, which gives the joint
probability density; in the present case, we have
$||\psi_{SN}(x_{1},x_{2})||^2 = ||\psi_S(x_{1})\psi_N(x_{2})||^2 +
||\psi_S(x_{2})\psi_N(x_{1})||^2 -
2\mathrm{Re}(\psi_S(x_{1})\psi_N(x_{2})\psi_S(x_{2})^{\ast}\psi_N(x_{1})^{\ast})$.
This last `interference term' (though vanishing at large distances),
cannot be dispensed with, and says that nothing, not even \emph{in
mente Dei}, can tell us which is the particular electron in the
South Pole (and the same happens for the North Pole). As far as
quantum physics is concerned, they really and truly have no identity
in the standard sense (and hence they have not \emph{identity} at
all).
\end{quote}

What is attacked in this quotation (and branded ``a common misunderstanding'') is the view  that a two-electron wave function of the form\footnote{We have inserted a factor $\frac{1}{\sqrt{2}}$ for the purpose of normalization.} 
\begin{equation}\psi_{SN}(x_{1},x_{2})= \frac{1}{\sqrt{2}}\{\psi_S(x_{1})\psi_N(x_{2})- \psi_N(x_{1})\psi_S(x_{2})\} \label{NS}
 \end{equation}
represents one electron located at the South Pole and one at the North Pole. The latter interpretation is certainly common, and it is also true that it conflicts with the Received View; but it should not for that reason be dismissed as a misunderstanding. On the contrary, it is part of a coherent and simple interpretation of (anti-)symmetric many-particle states; an interpretation with the advantages of not requiring the introduction of entities lacking identity and of being in accordance with physical practice.  

That it is consistent to interpret (anti-)symmetrized product states like the one in Eq.\ \ref{NS} as describing individual particles possessing distinguishing physical properties was pointed out by \cite{ghirardi}.\footnote{Perhaps the first to defend simillar ideas was \cite{lubberdink} in an unpublished master thesis.} This result can be used to develop an interpretation of bosonic and fermionic quantum states that is a rival to the Received View, as discussed in  \citep{die5,dieks10,diekslubb}, to which we refer. In the following we will summarize a number of key points. 

The objection put forward in the above quotation hinges on the fact that states of identical particles must be symmetric or anti-symmetric under permutations of the labels that occur in the state. As a consequence of this permutation symmetry the labels do not correspond uniquely to pure one-particle states: in the example of Eq.\ (\ref{NS}) neither the label ``$1$'' nor the label ``$2$'' belongs uniquely to the state $\psi_S$ or the state $\psi_N$---each of the two labels attaches equally to $\psi_S$ and $\psi_N$.  This motivates the statement, in the quotation, ``that nothing, not even \emph{in mente Dei}, can tell us which is the particular electron in the South Pole (and the same happens for the North Pole).'' The thought is that it is impossible to say whether particle $1$ or particle $2$ finds itself at any particular Pole. 

The silent premise of this argument is that the labels 1 and 2 are \emph{particle labels}: they refer to \emph{particle 1} and \emph{particle 2}, respectively. Since neither of these labels uniquely correlates to either the South or the North pole, in Eq.\ (\ref{NS}), it follows that the particles cannot be uniquely located at one of the poles.  

More generally, the (anti-)symmetry of states of particles of the same kind has the consequence that the labels are ``evenly distributed'' over all one-particle states occurring in the total state. The Received View considers this as proof that not even God could attribute a definite pure one-particle state to any of the particles.    

The interpretative maneuver that is able to dismantle this line of thought, and which we will defend here, is to associate particles \emph{not} with \emph{labels of  factor spaces}, but instead with the \textit{one-particle states} that occur in the total $N$-particle state.\footnote{The doctrine that the indices of the factor Hilbert spaces are also particle labels was dubbed ``factorism'' by \cite{caulton}. The position that we will defend here is therefore ``anti-factorist''. } In the example of Eq.\ (\ref{NS}) this means that we will consider the indices $1$ and $2$ as having a purely mathematical significance: they label the two factor Hilbert spaces of the total Hilbert space. By contrast, we will associate the two \emph{particles} with the two orthogonal states $\psi_S$ and $\psi_N$, respectively. This interpretative step eliminates the question of ``which particle is in which state'': the particle defined by $\psi_S$ is at the South Pole, the particle defined by $\psi_N$ at the North Pole.

In the case of fermions of the same kind (for example electrons) we have to work with anti-symmetric states, in which all one-particle states are mutually orthogonal and appear only once---this expresses  Pauli's exclusion principle. The particles defined by these orthogonal states are completely distinguishable at any instant of time (they are ``absolutely discernible'', \emph{i.e.}\ distinguishable by means of monadic physical properties \citep{die4}), since orthogonal quantum states can always be distinguished perfectly. This, then, bestows identity on fermions: distinguishability implies identity.  There \emph{may} also be identity over time, \emph{genidentity}. For example, this is the case if there is no interaction between $N$ fermions, so that each one-particle state evolves independently and unitarily. Orthogonal states remain orthogonal under unitary evolution, so that in this special case an $N$-fermion system corresponds to $N$ orthogonal one-particle states that trace out distinguishable paths in the total Hilbert space. 

Importantly, there is also approximate genidentity when the classical limit is approached. In this limit the quantum particles as defined in our rival to the Received View will gradually transform into their classical counterparts \citep{diekskrause}. 

However, in the most general quantum regime the existence of unrestricted genidentity cannot be guaranteed. Wave packets of particles defined at one instant of time will generally soon overlap with each other and be transformed by interactions, after which it may become undetermined which of the original particles is the same one as any given post-collision particle (see sections \ref{ident} and \ref{loss}). 

It may also happen that the total state is not an (anti-)symmetrized product, but rather a coherent superposition of product states. In such cases there will generally not be a simple  interpretation in terms of particles defined by pure one-particle states. This reflects a basic feature of the interpretation that we are discussing: it is not assumed that particles are fundamental in the quantum world. Rather, particles are \emph{emergent} entities \citep{die5,dieks10,diekslubb}. Only if certain conditions are fulfilled (\emph{e.g}., relating to decoherence or the classical limit) does a description in terms of particles become appropriate.   

Summing up, in the alternative to the Received View sketched here, particles are individuated by distinct physical properties, represented by mutually orthogonal one-particle quantum states; the factor Hilbert space indices that occur in the tensor product formalism of the quantum mechanics of identical particles are \emph{not} interpreted as particle labels. Quantum particles defined this way are always distinguishable. However, this notion of a quantum particle does not have unrestricted applicability: quantum particles are \emph{emergent}. This marks a contrast with the Received View: according to the Received View particles are fundamental building blocks of the world, even though their identity-less nature is mysterious. Our alternative says that particles  only emerge in particular physical circumstances; but in these cases they always possess their own identities.

\section{The Physics and Philosophy of Identical Particles With Identity}\label{ident}

In 1956, in a Letter to the Editor of the \emph{Physical Review}, the later Nobel prize winner Hans Dehmelt announced an experimental physics research program focusing on individual atoms and ions. \cite{dehmelt1956} predicted that for individual charged particles ``the intriguing possibility even exists to trap them by suitable fields''. As Dehmelt recalled in a 1990 review of his own work \citep{dehmelt1990}, it took another 17 years before he  finally succeeded in  confining a single electron quasipermanently in an electromagnetic trap \citep{dehmelt1973}. A decade later \cite{dehmelt1984} achieved a similar feat with a single positron, which was kept in a trap and observed continuously during three months---Dehmelt baptized this specific elementary particle ``Priscilla''. In 1989 Dehmelt, together with  Wolfgang Paul, was awarded the Nobel prize in physics ``for the development of the ion trap technique."

In 2012 the Nobel Prize for physics went to Serge Haroche and David Wineland ``for ground-breaking experimental methods that enable measuring and manipulation of individual quantum systems''---methods that elaborated on the work of Dehmelt. As the 2012 Nobel citation states  \citep{nobel2012}: ``The Nobel Laureates have opened the door to a new era of experimentation with quantum physics by demonstrating the direct observation of individual quantum particles without destroying them. ... David Wineland traps electrically charged atoms, or ions, controlling and measuring them with light, or photons. 
Serge Haroche takes the opposite approach: he controls and measures trapped photons, or particles of light, by sending atoms through a trap.'' Indeed, Haroche and his research group  had been able to study the behavior of a single photon that had been trapped in a cavity \citep{haroche}.

In their report the Nobel committee emphasized that these achievements should not be seen as engineering feats with little relevance for fundamental or philosophical questions.  As they wrote \citep[``Scientifc Background on the Nobel Prize in Physics'']{nobel2012}: 
\begin{quote}
These techniques have led to pioneering studies that test the basis of quantum mechanics and the transition between the microscopic and macroscopic worlds, not only in thought experiments but in reality. ... Wineland
and coworkers ... created [Schr\"odinger] ``cat states'' consisting of single trapped ions entangled with coherent states of motion and observed their decoherence. Recently, Haroche and coworkers created cat states, measured them and made a movie of how they evolve from a superposition of states to a classical mixture. ...
Today, the most advanced quantum computer technology is based on trapped ions, and has been demonstrated with up to 14 qubits and a series of gates and protocols.
\end{quote}

The remarks in the report about the transition from the quantum to the classical world are particularly pertinent to our present theme. The trapped particles are defined by their individual quantum states (well-localized in space, in the mentioned experiments) and can be counted in the ordinary sense. In the quantum computing case typical configurations consist of series of ``qubits'', positioned next to each other. These qubits can be distinguished, ordered and numbered, even though they are quantum systems of the same kind. Likewise, in the classical limit distinguishable quantum particles of the same kind mimic the behavior of classical particles; they thus realize the transition to the world of classical mechanics. This is all in accordance with the way in which our alternative to the Received View defines particles. By contrast, the particles of the Received View never possess any individuating characteristics and are therefore not the things the experiments described above focus on.

There are countless other examples in the physics literature in which identical quantum particles are treated as individual entities. That particles of the same kind can be dealt with as localized objects following more or less classical trajectories, if the spatial sizes of their wave packets are small compared to their mutual distances and if their momenta are relatively large, has been a standard assumption from the early days of quantum mechanics \citep{diekskrause}.\footnote{Note that this presupposes a characterization of particles via one-particle states, contrary to the Received View.} Proponents of the Received View certainly have a reason to try and undermine the validity of such practices. In this spirit Toraldo di Francia wrote \citep[p.\ 209]{toraldo}, \cite[p.\ 266]{toraldo2}:
\begin{quote}
...an engineer, discussing a drawing, can \emph{temporarily} make an exception to the anonymity principle and say for instance: `Electron $a$ issued from point $S$ will hit the screen at $P$ while electron $b$ issued from $T$ hits it at $Q$'. But this mock individuality of the particles has very brief duration.
\end{quote}

\cite{toraldo} did not explicate the term \emph{mock individuality}, but only made the brief comment that this ``individuality'' breaks down as soon as the electron encounters other electrons, for example by entering an atom. However, even if we set aside for the moment questions about what exactly happens in such processes, it remains obscure  why an individuality that does not last forever, perhaps even lasts only very briefly, should be dismissed as a ``mock individuality''.  After all, everyday macroscopic objects also have finite life spans, but it would seem weird to deny, on those grounds, their identity when they are still intact. 

In a later publication \cite{toraldo2} address this lacuna in their argumentation.  They start by admitting that particles could be defined, conventionally as they say, through their different states and thus even be given proper names (curiously, there is no reference to Dehmelt's Priscilla).  They also admit that the thus defined objects could last for a very long time. Still, they maintain, their \emph{possibly} limited life span raises a serious worry about the identity of such objects and the value of their names. They write \citep[p.\ 267]{toraldo2}:
\begin{quote}
\emph{Prima facie} one may be tempted to think that the case of particle $a$ is not different from that of Aristotle. After all, there was no Aristotle before 384 b.C.\ or after 322 b.C. But, in general, the particle does not die! An electron may very well survive a close encounter with another electron. Suppose that we follow with continuity an electron---say Peter---going from point $P$ to point $Q$ in a vacuum. We would like to be able to say that in a possible world Peter might encounter other electrons on its path and finally be scattered to $Q$. But then no one could tell that that electron is still Peter. There is no \emph{trans-world} identity. In this situation the meaning of `rigid designator' becomes very fuzzy. Anyway, the term seems useless.
\end{quote}

In order to judge this argument we need not enter into a discussion of possible worlds containing collisions that do not actually occur, the concept of trans-world identity, and the Kripkean notion of a ``rigid designator''. The physically relevant core of the above reasoning is that in a process in which electron Peter collides with another individual electron (Paul, say), after which there still are two individual electrons, it may be impossible to tell which of these post-collision electrons is Peter and which is Paul. This is deemed unacceptable by Dalla Chiara and Toraldo di Francia, since Peter has not died.
 
It is true that a loss of distinguishability of the described kind might occur---we will discuss this in the next section. But the argument as stated is unconvincing nonetheless.  If the identity of Peter and Paul were completely lost in their encounter\footnote{It is not absolutely necessary that such a loss occurs in an encounter, see section \ref{loss}.}, this would surely mean that both Peter and Paul have died as individual particles due to the collision, contrary to the premise in the above argument that particles in general do not die. It is as if Aristotle and his identical twin brother meet and coalesce, in a science fiction scenario, after which  two perfectly similar look-alikes reappear and start following their own courses. It is certainly possible to devise scenarios of this kind in which it makes no sense to ask who of the later two persons is the original Aristotle. But it does not follow  from this that it is futile to use the name ``Aristotle'' for the philosopher when he is still Aristotle, and that we should deny him his identity. The usefulness of giving names, and assigning identity, in analogous physics situations was exactly what was argued for in our above discussion of work leading to the 1989 and 2012 physics Nobel prizes.

D\'ecio Krause, one of the co-authors of the term ``Received View'' \citep{french} and one of the staunchest defenders of the View, has undertaken to investigate and criticize the claims of Dehmelt, Haroche and Wineland \citep{krause1,krause2}. Krause concludes that contrary to those claims, identical quantum particles always lack identity---even when they are trapped, are distinguishable and bear unambiguous names (like Priscilla). His argument relies on the quantum postulate that identical particles must be in permutation invariant states. Krause takes this postulate to imply that it does not make any difference to a physical situation when a particle, even a particle in a trap, is replaced by an arbitrary other particle of the same kind. This means, the argument continues, that in Dehmelt's experiment  there is no fact of the matter concerning the question \emph{which} positron in the universe is Priscilla. No positron has an identity, and Priscilla is only a ``mock name''.
   
Krause devotes a similar analysis to a situation with two trapped electrons, one in the infinite potential well (trap) $1$ and the other at some distance in a similar potential well $2$. The two-particle wave function for this case is anti-symmetric and has the form \citep{krause2}
\begin{equation}\label{psi12}
\psi_{12}(a,b) = \frac{1}{\sqrt{2}}\Big(\psi_1(a) \psi_2(b) - \psi_1(b) \psi_2(a)\Big),
\end{equation}
where $\psi_1$ and $\psi_2$ are wave functions that vanish everywhere except in well $1$ and well $2$, respectively; $a$ and $b$ are the labels of the two factor Hilbert spaces in the tensor product Hilbert space of the two-particle system. Krause comments:
\begin{quote}
note that we are dealing with different Hilbert spaces, the space of $a$ and the space of $b$. But, if the particles are indistinguishable, how can we know which particle is in well 1? In other terms, which particle has its states represented by vectors of the first space? There is no way to do it, for anyone of them could be in well 1.
\end{quote} 
The final conclusion must therefore be similar to the one drawn by Toraldo di Francia: 
\begin{quote}
Although trapped in the infinite wells, they [\emph{i.e., the quantum particles}] have only what Toraldo di Francia has termed mock individuality, an individuality (and, we could say, a `mock identity') that is lost as soon as the wells are open or when another similar particle is added to the well (if this was possible). And this of course cannot be associated with the idea of identity. Truly, there is no identity card for quantum particles.
 They are not individuals, yet can be isolated by trapping them for some time.
\end{quote}

Krause's argument accepts from the outset a key point of the Received View, namely that the labels occurring in the total state refer to \emph{particles}, and then focuses on the difficulty of answering the question \emph{which} particle is in the trap. The consistency with the Received View of the latter question is moot at best: according to the Received View all particles of the same kind have the same physical properties, so it cannot be the case that one of them finds itself in the trap while the others have different positions. But we do not want to press this point here---anyway, the argument does not represent a defense of the Received View, but merely an attempt to apply the ideas of that view to the case of the trapped particles.  As a defense it is question-begging.

Krause's reasoning fails from the start in our alternative to the Received View. According to this alternative, particles are characterized by the one-particle states occurring in the total state, instead of by labels. It is then  immediately clear that the state of Eq.\ (\ref{psi12}) represents two distinguishable particles, one in trap 1 and one in trap 2. 

Summing up, the attempt to apply ideas of the Received View to cases like that of Priscilla, the trapped positron, does not cast a favorable light on the View. Instead of providing a defense, it demonstrates the View's problematic aspects: even in cases where we have one particular electron in our hands, so to speak, we would have to accept that all electrons in the universe are utterly indistinguishable from each other.

\section{Loss of Distinguishability and Identity}\label{loss}

In 1987 Hong, Ou and Mandel performed a famous experiment that started a tradition of research in which individual photons, or atomic particles of the same kind, traveling along distinguishable paths, are brought together so that their wave functions overlap---this leads to a loss of distinguishability manifested by interference phenomena \citep{hom1,oumandel,hom2}. A recent version of the scheme, applied to two electrons, goes as follows \citep{tichy}. 

Suppose that we have two electron guns, one to the Left and one to the Right, and suppose that each of these devices fires exactly one electron---one with spin up in the $z$-direction, the other with spin down. (We follow the way in which experiments of this kind are commonly described in the physics literature, which is at odds with the Received View, as we have noted before.) Since electrons are identical fermions, the total wave function is anti-symmetric, so that it has the following form: 

\begin{equation}\label{bellvar}
    \frac{1}{\sqrt{2}} \{| L\rangle_1 | R \rangle_2 |\!\uparrow \rangle_1 |\! \downarrow \rangle_2 \ - | R \rangle_1 | L \rangle_2 |\!\downarrow \rangle_1 |\! \uparrow \rangle_2\}.
\end{equation}

The two individual  electron wave packets evolve independently, by free evolution; for the sake of simplicity we leave out this free part of the evolution. Now, after some time, each wave packet encounters a beam splitter and is split; this happens in such a way that, of both original packets, one half is directed to the location $L^\prime$ and the other half to the location $R^\prime$.

The effect of the two beam splitters can be represented in the following way:

\begin{eqnarray}\label{evolve1}
\ket{L}& \rightarrow & \frac{1}{\sqrt 2} \left( \ket{L^\prime}+\ket{R^\prime} \right),  \\ 
\ket{R}& \rightarrow & \frac{1}{\sqrt 2} \left( \ket{L^\prime}-\ket{R^\prime} \right)\label{evolve2} ,
\end{eqnarray}
where the states (ket vectors) $\ket{L^\prime}$ and $\ket{R^\prime}$ correspond to wave packets localized at $L^\prime$ and $R^\prime$, respectively.  

After the evolution, the total state is still an anti-symmetrized product. The two original orthogonal spatial states $\ket{L}$ and $\ket {R}$ have evolved into two new orthogonal states (the right-hand sides of (\ref{evolve1}) and (\ref{evolve2}))---let us call these $\phi$ and  $\psi$, respectively.  The final total state can now be written as 
\begin{equation}\label{bellvar1}
    \frac{1}{\sqrt{2}} \{| \phi \rangle_1 | \psi \rangle_2 |\!\uparrow \rangle_1 |\! \downarrow \rangle_2 \ - | \psi \rangle_1 | \phi
\rangle_2 |\!\downarrow \rangle_1 |\! \uparrow \rangle_2\}.
\end{equation}
 
According to our interpretation this state still represents two individual and distinguishable particles, defined by the orthogonal states $\ket{\phi} \ket{\!\!\uparrow}$ and $\ket{ \psi} \ket{\!\!\downarrow}$, respectively. This individual particle interpretation will be confirmed if we perform measurements of  observables like $\ket{\phi}\bra{\phi}$, $\ket{\psi}\bra{\psi}$, $\ket{\!\!\uparrow}\bra{ \uparrow\!\!}$ or $\ket{\!\downarrow}\bra{\downarrow\!}$; such measurements are able to distinguish perfectly between the two states defining our particles.

However, if we perform \emph{local} (in the spatial sense) spin measurements, by using electron spin detectors positioned at $L^\prime$ and $R^\prime$, an interpretation of the measurement results in terms of independent individual particles may appear problematic. This is so because \emph{both} of the two initial particles contribute to all detection results at  $L^\prime$ and  $R^\prime$, which leads to interference. A calculation shows \citep{tichy,dieks10} that correlations between spin values obtained at $L^\prime$ and $R^\prime$ will be found that suggest the presence of an entangled state (\emph{i.e.}, not a symmetrized product state) of the Einstein-Podolsky-Rosen type;  such a state has ``holistic'' characteristics and does not represent two independently existing particles each possessing its own set of distinctive physical properties \citep{die5,dieks10,diekslubb}.

This example illustrates how overlap of wave functions may veil the individuality of particles. In the discussed case there were always two individual particles, also during their encounter. Each kept its own identity, and this could have been verified by performing appropriate measurements. But if only the results of local measurements in the overlap area are available this will suggest a loss of individuality.

These comments address part of the worries expressed by \cite{toraldo,toraldo2,krause1,krause2}. They illustrate that identity grounded in individual and orthogonal one-particle states is not automatically and immediately lost outside of traps and in ``close encounters''.  

It remains true, however, that during interactions (anti-)symmetric product states may be converted into superpositions of such states, which are not products themselves. In such cases entangled states are created and  there will consequently be a loss of identity, because it will no longer be possible to decompose the total state into one-particle states.  

In the case of bosons there is the additional possibility that two originally orthogonal states, both with occupation number $1$, will evolve into a new state with occupation number $2$, and similarly for more bosons. This would also imply a loss of identity: two individual particles would merge into one object, one whole, which is no longer an elementary particle.    

In such situations the applicability of the concept of an individual elementary particle will  end. The original particle will ``die'', and its identity will die with it. However, as argued in section \ref{ident}, a finite life span does not degrade identity into mock identity.

\section{Conclusion}

According to the Received View the world described by quantum mechanics consists of elementary building blocks that lack identity and share all their physical properties. This picture is hugely different from what is suggested by the actual practice of quantum physics and by the limiting cases in which classical physics becomes applicable. We have sketched an alternative to the Received View, drawing on earlier work, and compared and contrasted it with its rival. According to this alternative, systems of identical quantum particles can in some cases be described as consisting of individuals, possessing their own identity. This view makes sense of the way such systems are usually discussed in the practice of physics, and it also provides an understandable and simple story of the way in which the classical world emerges from the quantum realm.   

Differently from the Received View, the alternative does not maintain that the world on its deepest quantum level is always particle-like. Particles \emph{emerge}, under certain conditions. The typical quantum phenomenon of a Bose-Einstein condensate provides an example of the difference between the two approaches. According to the Received View such a condensate should be conceptualized as consisting of many bosons, though indistinguishable and lacking identity. This is analogous to thinking of a bank account in terms of different euros, existing independently of each other but without identity. The alternative view adopted here is that a bank account has a certain value, a purchasing power, that is not composed of independent constituents. Similarly, a Bose-Einstein condensate is one object, having a total mass and charge without independent components. But when this object is subjected to certain interactions or measurements, individual bosons may emerge. In the case of fermions, one-particle states can only be singly occupied, and this makes it easier for fermions than for bosons to manifest themselves as individual particles. 

We believe that this alternative to the Received View not only fits physical practice better, but also provides a consistent and understandable perspective that improves on the obscure notion of identity-less physical objects.


\end{document}